\documentclass[longauth,traditabstract]{aa} 
\usepackage{graphicx}
\usepackage{equationarray}
\usepackage{txfonts}
\usepackage{natbib}
\bibpunct{(}{)}{;}{a}{}{,} 
\newcommand{\nc}{\newcommand}
\nc{\lsun}{\ensuremath{\mathrm{L}_\odot}}
\nc{\msun}{\ensuremath{\mathrm{M}_\odot}}
\nc{\tex}{\ensuremath{\mathrm{T}_{\rm ex}}}
\nc{\cthree}{C$_3$}
\nc{\kms}{\mbox{km\,s$^{-1}$}}
\nc{\Kkms}{\mbox{K\,km\,s$^{-1}$}}
\nc\micron{\mbox{$\mu$m}}
\nc{\Trot}{$T_{\rm rot}$}%
\nc{\Ntot}{$N(C_3)$}%
\nc{\Tc}{$T_{\rm c}$}%
\nc{\Tdust}{$T_{\rm dust}$}%
\nc{\Tex}{$T_{\rm ex}$}%
\nc{\Tkin}{$T_{\rm kin}$}%
\nc{\cmcub}{\mbox{cm$^{-3}$}}
\nc{\cmsq}{\mbox{cm$^{-2}$}}

\begin{document}
   \title{
Excitation and Abundance of C$_3$ in star forming cores:}
    \subtitle{
    Herschel/HIFI \thanks{Herschel is an ESA space observatory
    with science instruments provided by European-led Principal
    Investigator consortia and with important participation from NASA.}
    observations of the sight-lines to W31C and W49N
}

   \author{
B.~Mookerjea\inst{\ref{tifr}}, 
T.~Giesen\inst{\ref{kosma}}, 
J.~Stutzki\inst{\ref{kosma}}, 
J.~Cernicharo\inst{\ref{cantab}},  
J.~R.~Goicoechea \inst{\ref{cantab}}, 
M.~De Luca\inst{\ref{ens}},
T.~A.~Bell\inst{\ref{caltech}}, H.~Gupta\inst{\ref{jpl}}, M.~Gerin\inst{\ref{ens}}, 
C.~M.~Persson
\inst{\ref{chalmers}},  P.~Sonnentrucker\inst{\ref{jhu}},
Z.~Makai \inst{\ref{kosma}},
J.~Black\inst{\ref{chalmers}}, F.~Boulanger\inst{\ref{ias}},
A.~Coutens\inst{\ref{toulouse}}, E.~Dartois
\inst{\ref{ias}}, P.~Encrenaz
\inst{\ref{ens}}, E.~Falgarone \inst{\ref{ens}},  
 T.~Geballe\inst{\ref{gemini}},
B.~Godard\inst{\ref{ens}},P.~F.~Goldsmith\inst{\ref{jpl}}, 
C.~Gry \inst{\ref{marseilles}}, P.~Hennebelle
\inst{\ref{ens}}, E.~Herbst\inst{\ref{ohio}}, 
P.~Hily-Blant \inst{\ref{grenoble}}, C.~Joblin
\inst{\ref{toulouse}}, M.~Ka{\'z}mierczak\inst{\ref{torun}}, 
R.~Ko{\l}os  \inst{\ref{chempoland}}, 
J.~Kre{\l}owski\inst{\ref{torun}}, 
D.~C.~Lis \inst{\ref{caltech}}, J.~Martin-Pintado \inst{\ref{cantab}},
K.~M.~Menten \inst{\ref{mpifr}}, 
R.~Monje \inst{\ref{caltech}},
J.~C.~Pearson\inst{\ref{jpl}}, M.~Perault \inst{\ref{ens}}, 
T.~G.~Phillips \inst{\ref{caltech}}, 
R.~Plume \inst{\ref{calgary}},  M.~Salez \inst{\ref{ens}}, 
S.~Schlemmer \inst{\ref{kosma}}, 
M.~Schmidt\inst{\ref{ncacpoland}},
D.~Teyssier\inst{\ref{esamadrid}}, C.~Vastel
\inst{\ref{toulouse}}, 
S.~Yu \inst{\ref{ens}}, 
P. Dieleman\inst{\ref{sron}}, 
R. G\"usten\inst{\ref{mpifr}}, 
C.~E.~Honingh\inst{\ref{kosma}}, 
P.~Morris\inst{\ref{ipac}}, 
P.~Roelfsema\inst{\ref{sron}}, 
R.~Schieder\inst{\ref{kosma}}, 
A.~G.~G.~M. Tielens\inst{\ref{leiden}}
\and J.~Zmuidzinas\inst{\ref{caltech}}
}

\institute{Tata Institute of Fundamental Research, Homi Bhabha Road,
Mumbai 400005, India \email{bhaswati@tifr.res.in}\label{tifr} 
\and 
I. Physikalisches Institut, University of Cologne, Germany\label{kosma}
\and
Centro de Astrobiolog\'{\i}a, CSIC-INTA, 28850, Madrid, Spain\label{cantab}
\and LERMA, CNRS, Observatoire de Paris and ENS, France \label{ens}
\and California Institute of Technology, Pasadena, CA 91125, USA\label{caltech}
\and JPL, California Institute of Technology, Pasadena, USA\label{jpl}
\and  Onsala Space Observatory, Chalmers University of Technology, SE-43992 Onsala, Sweden \label{chalmers} 
\and Depts.\ of Physics, Astronomy \& Chemistry, Ohio State Univ. USA.\label{ohio}  
\and The Johns Hopkins University, Baltimore, MD 21218, USA\label{jhu}
\and Institut d'Astrophysique Spatiale (IAS), Orsay, France.\label{ias} 
\and Laboratoire d'Astrophysique de Marseille (LAM), France.\label{marseilles} 
\and Laboratoire d'Astrophysique de Grenoble, France.\label{grenoble}
\and Universit\'e Toulouse; UPS ; CESR ; and CNRS ; UMR5187,
9 avenue du colonel Roche, F-31028 Toulouse cedex 4, France
\label{toulouse}
\and Gemini telescope, Hilo, Hawaii, USA \label{gemini}.
\and MPI f\"ur Radioastronomie, Bonn, Germany \label{mpifr}.
\and Institute of Physical Chemistry, PAS, Warsaw, Poland\label{chempoland}
\and Nicolaus Copernicus University, Toru{\'n}, Poland\label{torun}
\and Dept.\ of Physics \& Astronomy, University of Calgary, Canada\label{calgary}
\and  Nicolaus Copernicus Astronomical Center (CMAK), Toru{\'n}, Poland\label{ncacpoland}
\and European Space Astronomy Centre, ESA, Madrid, Spain\label{esamadrid} 
\and  Infrared Processing and Analysis Center, California Institute of Technology, MS 100-22, Pasadena, CA 91125\label{ipac}
\and  Sterrewacht Leiden, University of Leiden, Leiden, The Netherlands\label{leiden}
\and SRON Netherlands Institute for Space Research, Landleven 12, 9747 AD Groningen, The Netherlands\label{sron}
}

  \date{Received \ldots accepted \ldots}

\abstract
{We present spectrally resolved observations of triatomic carbon
(\cthree) in several ro-vibrational transitions between the
vibrational ground state and the low-energy $\nu_2$ bending mode  at
frequencies between 1654--1897\,GHz along the sight-lines to the
submillimeter continuum sources W31C and W49N, using {\em Herschel's}
HIFI instrument.  We detect \cthree\ in absorption arising from the
warm envelope surrounding the hot core, as indicated by the velocity
peak position and shape of the line profile.  The sensitivity does not
allow to detect \cthree\ absorption due to diffuse foreground clouds.
From the column densities of the rotational levels in the vibrational
ground state probed by the absorption we derive a rotation temperature
(\Trot) of $\sim$\,50--70 ~K, which is a good measure of the
kinetic temperature of the absorbing gas, as radiative transitions
within the vibrational ground state are forbidden. It is also in good
agreement with the dust temperatures for  W31C and W49N.  Applying the
partition function correction based on the derived \Trot, we get
column densities  N(\cthree)$\sim$ 7--9$\times10^{14}$~\cmsq\
and abundance x(\cthree)$\sim$$10^{-8}$ with respect to H$_2$. For
W31C, using a radiative transfer model including far-infrared pumping
by the dust continuum and a temperature gradient within the source
along the line of sight we find that a model with
x(\cthree)=10$^{-8}$, \Tkin=30-50 K, N(\cthree)=1.5 10$^{15}$
cm$^{-2}$ fits the observations reasonably well and provides
parameters in very good agreement with the simple excitation analysis.
}

\keywords{ISM:~molecules -- Submillimeter:~ISM -- ISM:lines and bands
-- ISM:individual (W49N, W31C) --line:identification -- molecular data
-- Radiative transfer  }

  \titlerunning{Detection of \cthree\ towards W31C and W49N}
        \authorrunning{Mookerjea et al.}
   \maketitle

\section{Introduction}

Small carbon chains are relevant in the chemistry of stellar and
interstellar environments for several reasons: 
ubiquitous interstellar spatial distribution, they likely participate
in the formation of long carbon chain molecules, and they are products
in photo-fragmentation processes of larger species such as PAHs.
Triatomic carbon, \cthree, was first tentatively identified in
interstellar gas by \citet{vanorden1995} and \citet{haffner1995}.  The
mid-infrared spectrum of \cthree\ ($\nu_3$ antisymmetric stretching
mode) was measured in the circumstellar envelope of CW Leo (IRC
+10216) by \citet{hinkle1988}, and in low-resolution interstellar
absorption  in the far-IR ($\nu_2$ bending mode) toward Sgr B2 by
\citet{cernicharo2000}.  \citet{giesen2001} discussed new laboratory
data on the vibrational spectrum of \cthree\ in its low-frequency
bending mode and re-visited the first identification of the $\nu_2$
$R(2)$ line in absorption toward Sgr B2 \citep{vanorden1995}. The
abundance and excitation of \cthree\ in translucent clouds were
determined convincingly by \citet{maier2001}, \citet{roueff2002} and
\citet{oka2003} at optical wavelengths.

The Heterodyne Instrument for the Far-Infrared
\citep[HIFI;][]{deGraauw2010} on board the Herschel Space Observatory
\citep{pilbratt2010}, with its broad frequency coverage, high
sensitivity and spectral resolution provides for the first time the
opportunity for a systematic study of carbon chain molecules such as
\cthree\ through probing several ro-vibrational lines at full spectral
resolution.  In this Letter, we present the first results of our
search for \cthree\ obtained from observations of the sight-lines to
the bright far-infrared (FIR) continuum sources W31C  and W49N as part
of the PRISMAS (``PRobing InterStellar Molecules with Absorption line
Studies”) Key Program  \citep{gerin2010}.

W31C (G$10.6-0.4$) is one of three bright HII regions within the W31
complex, and an extremely luminous submillimeter and infrared
continuum source \citep[$L_{\rm IR} \sim 10^7$\,\lsun][]{wright1977}.
Located at a distance of $4.8^{+0.4}_{-0.8}$~kpc \citep{fish2003} with
v$_{\rm LSR}$ = -3 to -1 \kms\ \citep{miettinen2006}, the sight-line
to G$10.6-0.4$ intersects several foreground molecular clouds
\citep{corbel2004}. W49N is one of the three main IR peaks of W49A,
which has a luminosity of \citep[$L_{\rm
bol}\sim10^7$~\lsun][]{wardthompson1990}. It is located at a distance
of 11.4\,kpc and has v$_{\rm LSR}$ = 12~\kms. The sight-line toward W49N
is spectroscopically interesting because of the numerous features
contributed by W49A itself, as well as by additional clouds associated
with the Sagittarius spiral arm (which crosses the line of sight
twice). Numerous spectroscopic studies (both emission and absorption)
have been carried out in the past towards both these sources
\citep[e.g.,][]{plume2004,neufeld2010}.

\section{ C$_3$ energy level diagram and radiative transitions }
\begin{figure}
\begin{center}
\includegraphics[width=0.6\textwidth]{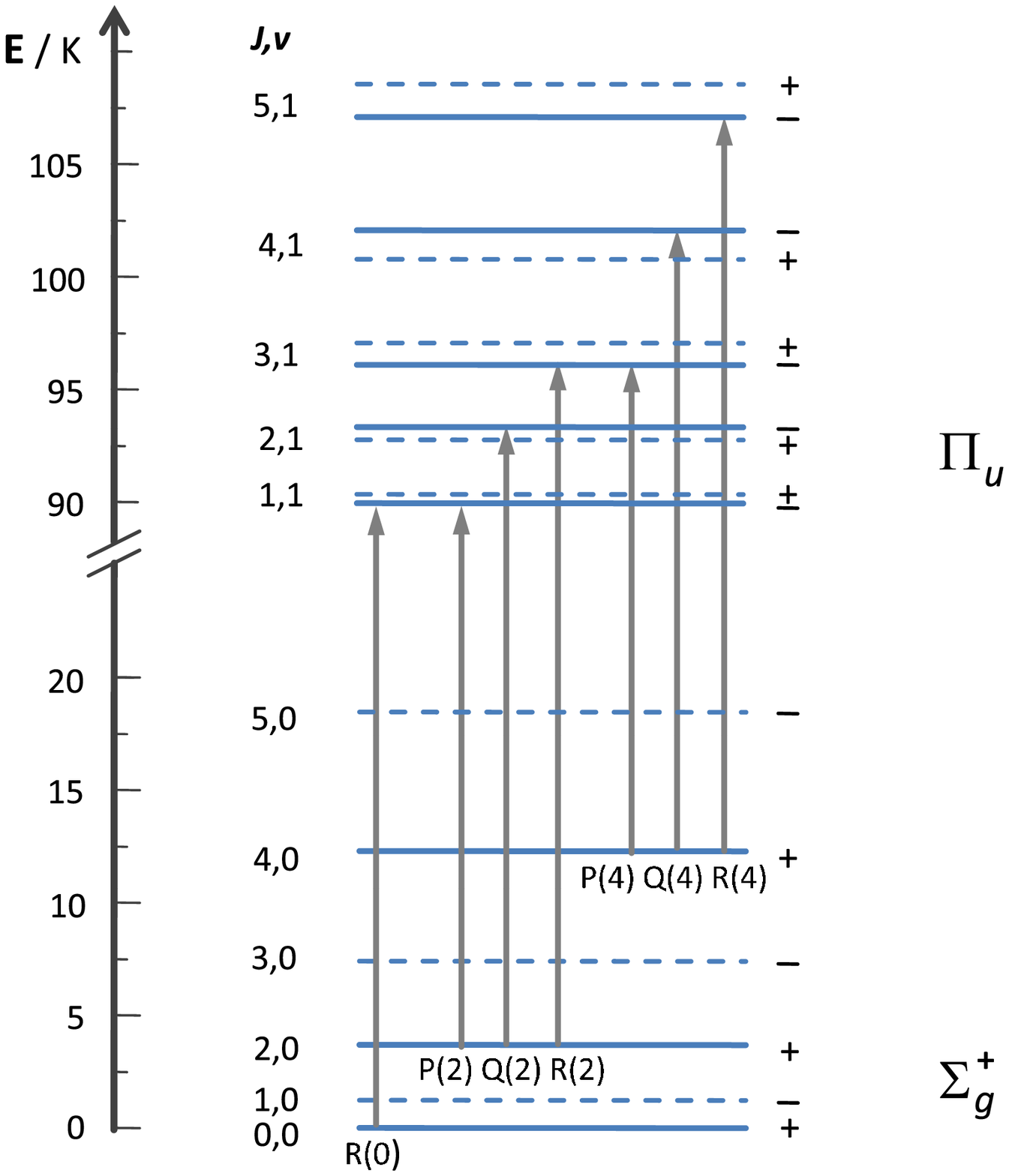}
\caption{Energy level diagram of
 $\Sigma_g^+$  ground state and $\Pi_u$ lowest  bending mode state of \cthree.
Due to nuclear spin statistics half of the rotational energy levels (dashed
lines) are missing. Allowed ro-vibrational $P-$, $Q-$, and $R-$branch
transitions from $v=0$ to $v^\prime =1$  follow $- \leftarrow +$ selection rules.
\label{fig_c3energy}}
\end{center}
\end{figure}

Linear C$_3$ has an energetically unusually low $\nu_2$ bending mode
at only $\sim90$ K.  Ro-vibrational transitions of the $\nu_2$ band in
its $^1\Sigma^+_g$ electronic ground state have been reported by
\citet{gendriesch2003} and \citet{schmuttenmaer1990}.  $P(J)$
transitions with $J=2,4,6$, $R(J)$ with $J=0,2,4,6$, and $Q(J)$ with
$J=2,4,$\ldots 16 have been measured in the laboratory, using high
resolution Terahertz sideband spectrometers at Berkeley
\citep{schmuttenmaer1990} and Cologne \citep{gendriesch2003} with
frequency accuracies of 7 MHz and 0.5 MHz respectively, which
corresponds to line frequency uncertainties of 1.1 \kms\ and 0.08
\kms.  All data were used in a global fit analysis to obtain most
accurate molecular constants \citep{giesen2001} which are presented as
line lists for astronomical observations in the Cologne Data Base for
Molecular Spectroscopy (CDMS) \citep{mueller2005}.  The ground
state of C$_3$ has $^1\Sigma^+_g$ symmetry while the vibrational
excited $\nu_2$ is a two-fold degenerate bending state of $\Pi_u$
symmetry.  Due to the $^{12}$C nuclear spin of $I=0$ in the ground
state only levels of (+) parity are present, while for the excited
bending state only levels with (-) parity are allowed.  As a
consequence, in the ground state of C$_3$ all odd numbered $J$
rotational levels are missing, whereas in the vibrational excited
$\Pi_u$ state both, even and odd $J$ rotational levels are present,
but the two-fold degeneracy of the vibrational state is lifted (see
Fig.~\ref{fig_c3energy}). Consequently, the statistical weights of the
ro-vibrational levels is simply given by the rotational degeneracy:
$g_{J,v}=2\,J+1$. The $\nu_2$ bending mode has a perpendicular type
spectrum with a calculated vibrational dipole moment of 0.437 Debye
\citep{jensen1992} which shows prominent $Q-$, $P-$, and $R-$branch
transitions.

For the analysis of the HIFI \cthree\ data, rest frequencies for $P(4)$,
$Q(2)$, and $Q(4)$ were taken from laboratory measurements while for
$P(10)$ the rest frequency has been obtained from a global fit of all
available laboratory data.

\begin{table}
\begin{center}
\caption{Spectroscopic parameters for the observed \cthree\
transitions \label{tab_labdata}}
\begin{tabular}{p{0.5cm}r@{$\leftarrow$}lr@{.}lcrr}
\hline \hline
Name &
\multicolumn{2}{c}{Transition}   &
\multicolumn{2}{c}{Frequency $^a$} & 
{A-coeff} & E$_l$\\
\ & ($J^\prime$,$v^\prime$)&($J,v$)
& \multicolumn{2}{c}{[MHz]} & $10^{-3}$s$^{-1}$ &  [K]\\
\hline
$P(10)$ & (9,1) & (10,0)& 1654081&66(4.68) $^b$  &  2.38 & 47.3\\
$P(4)$  & (3,1) & (4,0) &1787890&57(6.90)  &  2.72 & 8.6 \\
$Q(2)$  & (2,1) & (2,0) &1890558&06(0.25)  &   7.51  & 2.6 \\
$Q(4)$  & (4,1) & (4,0) &1896706&56(0.15)  & 7.58 &   8.6  \\
\hline \hline
\end{tabular}
\end{center}

$^a$ Experimental rest frequencies, uncertainties are given in parentheses.\\
 $^b$  Calculated frequency and 1$\sigma$ uncertainty taken from CDMS catalog.
\\
\end{table}

\section{Observations and data reduction}

Along the sight-lines to W31C and W49N we have observed four lines of
the $\nu_2$ bending mode, $P(4)$, $Q(2)$, $Q(4)$ and $P(10)$ (the
latter only in W31C), of triatomic carbon in the upper sideband of the
HIFI bands 7a, 7b and the lower sideband of band 6b of the HIFI
receiver.   The observations of W31C and W49N were carried out on 2010
March 8 and 2010 April 19 respectively.  The $P(10)$ line was
available as a ``bonus" for an LO tuning dedicated  to observe the CH
line at 1661\,GHz and it is yet to be observed in W49N with HIFI. All
observations are in dual beam switch (DBS) mode and with the Wide Band
Spectrometer with its spectral resolution of 1.1 MHz, corresponding to
a velocity resolution of $\sim$ 0.17 km~s$^{-1}$ at the frequencies of
the \cthree\ lines.  To identify the line origin from the lower and
upper sidebands, each line was observed with three LO settings shifted
by 15 km~s$^{-1}$ relative to each other.  The $Q(2)$ line also shows
up in the $Q(4)$ observations from the lower sideband, and, for one of the
LO tunings, partially overlap with the former.  The data were
first processed with HIPE \citep{ott2010}, and subsequently exported
to CLASS. At the high frequencies for these observations the H and V
polarizations were at times found to be discrepant in the measured
continuum level.  Observations optimized for reliable continuum
measurement were used to select the spectra with the correct continuum
level, used for the subsequent analysis.  All spectra were smoothed to
a resolution of $\sim$ 0.68~\kms\ and the rms noise level for the
spectra lie between 0.01--0.03~K.  Table~\ref{tab_gaussfit} gives the
measured double sideband continuum level (\Tc).  For the remainder of
the paper we discuss the line intensities normalized to the
single-sideband continuum level, where we have assumed a sideband gain
ratio of unity. 

\section{Results}

\begin{figure}
\begin{center}
\includegraphics[width=0.4\textwidth]{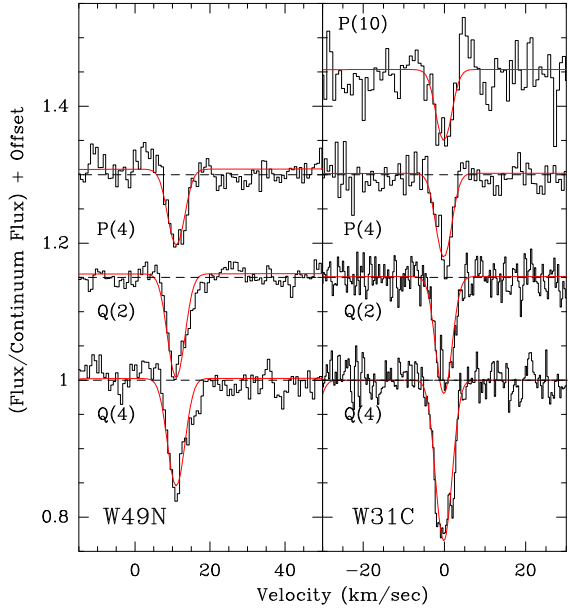}
\caption{The observed absorption spectra of the $P(4)$, $P(10)$, $Q(2)$ and $Q(4)$
ro-vibrational transitions of C$_3$ towards W49N ({\em left}) panel
and W31C ({\em right} panel), plotted along with 
the best-fit profiles 
obtained by simultaneous Gaussian fitting to all lines.  
\label{fig_specs}} 
\end{center}
\end{figure}

We have for the first time detected the spectrally resolved $\nu_2$
band transitions $P(4)$, $P(10)$, $Q(2)$ and $Q(4)$ lines of \cthree.
Fig.~\ref{fig_specs} shows the observed spectra normalized to the
(single sideband) continuum level.  A multi-component Gaussian with
common velocity width and spacing and individual amplitudes for each
line was fitted to derive the basic parameters of the absorption
spectra.  Table~\ref{tab_gaussfit} presents the fit results and their
uncertainties.  The $P(10)$ spectrum is affected by the spectral lines
of CH (1661.10726\,GHz) and H$_2$O (1661.007637\,GHz) from the upper
sideband. 

The \cthree\ absorption features towards W31C and W49N are centered
near 0\,\kms\ and 11\,\kms\ respectively.  The systemic velocities of
W31C and W49N are at $\rm v_{\rm LSR}$ = $-1$~\kms\ and 12~\kms.   Thus
the \cthree\ absorption lines detected here appear to be physically
associated with the hot core itself and most likely arise in the lower
density warm envelope surrounding them.  We do not detect any
absorption feature arising from the foreground clouds towards either of
these sources (see Sec.~\ref{sec_discussion}).

\begin{table}
\begin{center}
\caption{Parameters derived from simultaneous Gaussian fitting to all
the line profiles and column densities estimated from the fitted
intensities.  The fitted values of line center and linewidth are: for
W31C $V_{\rm cen}$= -0.09$\pm$0.06\,\kms\ and $\Delta V$ =
4.5$\pm$0.1\,\kms\ and  for W49N $V_{\rm cen}$=11.0$\pm$0.1\,\kms, and
$\Delta V$ = 5.3$\pm$0.2\,\kms.
\label{tab_gaussfit}}
\tiny
\begin{tabular}{l|lll|lll}
\hline
\hline
\multicolumn{1}{c|}{Transition} &
\multicolumn{3}{|c|}{W31C} &
\multicolumn{3}{|c}{W49N} \\
&\hspace*{0.3cm}\Tc & $\int{\tau d\rm v}$ & \hspace*{0.2cm}N$_{\rm l}$
&\hspace*{0.3cm}\Tc & $\int{\tau d\rm v}$ & \hspace*{0.2cm}N$_{\rm l}$\\
& \hspace*{0.3cm}[K] & 
[km/s] 
&  [\cmsq] &\hspace*{0.3cm}[K] &  
[km/s] 
& [\cmsq]\\
\hline
$P(10)$& 11.4$\pm$0.4 & 0.51 & 1.0\,10$^{14}$ & \ldots       &  \ldots \ldots\\
$P(4)$ & 10.5$\pm$0.3 & 0.61 & 1.5\,10$^{14}$ & 16.8$\pm$0.2 & 0.62 & 1.6\,10$^{14}$  \\
$Q(2)$ & 11.8$\pm$0.2 & 0.87 & 7.2\,10$^{13}$ & 18.3$\pm$0.3 &  0.87  & 7.3\,10$^{13}$\\
$Q(4)$ & 11.8$\pm$0.2 & 1.22 & 1.0\,10$^{14}$ & 18.3$\pm$0.2 & 0.90 &  7.6\,10$^{13}$\\
\hline
\end{tabular}
\end{center}
\end{table}

\subsection{Two-layer excitation analysis}

We first consider a simple two-layer model in which a (warm) absorbing
layer without continuum opacity lies in front of a (hotter) emitting
background source.  We used the formalism explained in detail in
Appendix A to obtain estimates of the rotational temperature (\Trot)
from the state specific column densities (Tab.~\ref{tab_gaussfit}).
The $J=4$ column densities are redundantly determined through both the
$P(4)$ and $Q(4)$ transitions.  One of the primary sources of error in
determination of the state specific column densities is the
determination of the continuum levels. We estimate the uncertainty in
the derived column densities to be 20\%.  For both W31C and W49N
the discrepancy in the $J=4$ column densities determined from $P(4)$ and
$Q(4)$ is much larger than this uncertainty.  We can only speculate on
the reason: firstly, the assumptions of the simple two-layer model may
not be appropriate and a more sophisticated model should be used (see
below). Secondly, we note that in case the \cthree\ absorption fills
only a fraction of the solid angle of the continuum source within the
beam, the intrinsic absorption would be much larger, possibly reaching
saturation for the stronger $Q(4)$ line.  We use an average of the
column densities at $J=4$ obtained from $P(4)$ and $Q(4)$ for the rest
of the paper. 

Using the formulae from Appendix A, we obtain for W49N a \Trot\
of 70\,K with a large formal error. This is consistent with \Tdust\
derived from dust continuum observations as 43\,K \citep{vastel2000}
since $T_{\rm gas}$ can be higher than \Tdust\ through direct gas
heating near the core.  For \Trot = 70\,K we calculate from the $J$=2
column density a total column density of N(\cthree) =
9\,10$^{14}$~\cmsq.

For W31C we have used the state-specific column densities ($N_J$) at
$J$=2, 4 and 10, to determine \Trot\ and \Ntot\ from the slope and the
intercept respectively of a linear fit to the log($\frac{N_J}{2J+1}$)
vs $E_J$ points (``rotation diagram").  Assuming a uniform 20\%  error
in the determination of the column densities we obtain \Trot =
56$\pm$6\,K and \Ntot = (7$\pm$0.5)\,10$^{14}$\,\cmsq. The estimated
\Trot\ is consistent with the observed dust temperature of 52\,K
derived from continuum observations by \citet{mueller2002}.

\subsection{Radiative transfer models for \cthree\ excitation in W31C}

\begin{figure}
\begin{center}
\includegraphics[width=0.25\textwidth,angle=-90]{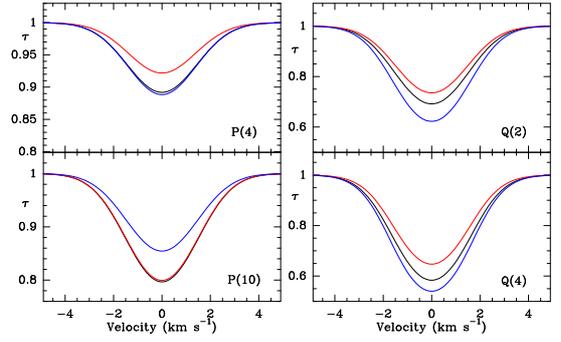}
\caption{Model predictions for the $Q(2)$, $Q(4)$, $P(2)$ and
$P(10)$ lines of the $\nu_2$=1--0 transition of \cthree\ in W31C. The black
lines correspond to a model with $n$(H$_2$)=10$^5$ cm$^{-3}$,
$x$(\cthree)=5 10$^{-8}$ and \Tkin=50 K. The red lines correspond
to a model with  $n$(H$_2$)=5 10$^5$ cm$^{-3}$,
$x$(\cthree)=10$^{-8}$ and \Tkin=50 K. Finally, the blue lines
correspond to the same parameters as the red ones except
that \Tkin=30 K.
\label{fig_w31cmodel}}
\end{center}
\end{figure}

In the likely scenario that the dust continuum source and its
associated continuum opacity are coexistent with the gas that absorbs
in the \cthree\ lines, the \cthree\ molecule will be embedded in a
relatively strong continuum radiation which would contribute to the
ro-vibrational excitation. In addition, the source intrinsic continuum
will partially fill in the line absorption. Thus, in a more detailed
approach we use a radiative transfer model, which considers FIR pumping
by the dust continuum \citep{cernicharo2000} and a temperature
gradient of the continuum source along the line of sight. For W31C we
find that the C$_3$ column densities can be interpreted by a cloud
which is twice the size of the continuum source, has a molecular
hydrogen density $n$(H$_2$) = 10$^5$ cm$^{-3}$, v$_{\rm turb}$= 2
\kms, abundance $x$(C$_3$) = 5 10$^{-8}$,  kinetic temperature of 50
K and  \Ntot = 1.5 10$^{15}$ cm$^{-2}$ (see
Fig.~\ref{fig_w31cmodel}).  The fact that the \cthree\ column density
comes out larger in this more detailed model is expected, as the dust
opacity partially fills-in the absorption line.

The main source of uncertainty in these models are the ro-vibrational
collisional rates which are based on  rather crude assumptions
\citep{cernicharo2000}. However, irrespective of the adopted collisional
rates the ground state is always thermalized, even in the presence of a
strong IR radiation field in our models. We find that owing to the lack
of a permanent electrical dipole moment the uncertainties in the
collisional rates have little effect on the emerging intensities.
Fig.~\ref{fig_w31cmodel} also shows that the resulting absorption depths
are almost unaffected by a change in the local density (and hence
abundance) by as much as a factor of 5, as long as the column density
remains constant.  We find that the $\nu_2$=1--0 transitions are
dominated by infrared pumping. The ro-vibrational excitation temperature
in the inner part of the cloud is 35--37 K for the lowest rotational
levels and around 20 for the higher-J states.  In the external layers
$T_ {\rm ex}$ decreases by 20 and 10 K, respectively. In a second model with
$n$(H$_2$) = 5 10$^5$ cm$^{-3}$ and $x$(C$_3$)=10$^{-8}$ the results are
almost equally consistent with the observations.  We also note, that
depending on the geometrical arrangement, the FIR-pumping in the
ro-vibrational transitions can result in the net effect of lowering the
rotational temperature in the vibrational ground state slightly below
the kinetic temperature of the gas.

We find that the major effect on the resulting absorption depths is
related to the kinetic temperature adopted for the absorbing gas.  In
Fig.~\ref{fig_w31cmodel} the blue lines correspond to a gas with
\Tkin=30 K. The high-$J$ lines of the ground state are less populated
than in the previous case and the opacities for the $\nu_2=$1--0
transitions decrease. If the opacity of the ro-vibrational lines
become larger than 1, and the central continuum source is optically
thick at the wavelengths of the $\nu_2$=1--0 transitions, then the
$\nu_2$ level becomes thermalized to the temperature of the dust.
This effect can be counterbalanced by decreasing the gas temperature
in the external layers of the cloud.  Clearly, a more elaborate
analysis than can be presented in this letter, is needed to explore
the full parameter range.

\section{Discussion
\label{sec_discussion}}

In the absence of allowed radiative transitions in the ground state,
the excitation of \cthree\ in the $\nu_2=0$ state can be assumed to 
be thermalized to the kinetic
temperature: \Trot = \Tkin. However, \Trot\ can be 
larger than \Tdust\ in the presence of direct gas heating mechanism
like the photoelectric heating.  The $\nu_2$ mode on the other hand
could be excited by collisions and as well as by infrared photons.
With Einstein coefficients ranging between 2 and
7~10$^{-3}$\,s$^{-1}$, the line opacities can be high and the infrared
pumping rather efficient. As a result, the excitation temperatures 
of the ro-vibrational lines are much
lower, typically between the beam-diluted \Tdust\ and \Tkin\ and hence
the lines are seen in absorption.  However, as explained above, the
excitation temperature within the rotation ladder of the ground
vibrational state can have values larger than \Tdust.

Crude estimates based on the ``two-layer" approach derive a \Trot\
consistent with the dust temperature of the continuum source and give
column densities of \cthree\ in W31C and W49N to be between 
7--9$\times 10^{14}$~\cmsq. Using the more elaborate radiative transfer
model we derived a column density of 1.5~10$^{15}$~\cmsq\ for W31C for
\Tkin\ of 30\,K.  Based on the discussion above we further argue
that the thus derived column density of \cthree\ is likely to be a
lower limit.  Implicit assumptions like a source filling factor of
unity, \Tex $\ll$ \Tc\ etc.  in the two-layer approach translate to an
underestimate of \Trot.  Moreover the source intrinsic continuum
opacity at 1.9~THz partially re-fills the absorption (as shown in the
radiative transfer model), and hence the column densities derived from
the radiative transfer model are higher than those derived in the
two-layer model.  Thus taking all uncertainties into account we
conclude that for both W31C and W49N the \cthree\ column densities are
$\sim 10^{15}$~\cmsq, correct to within a factor of 2 or so.  The
H$_2$ column densities in W31C and W49N are $9.2\times 10^{22}$~\cmsq\
\citep{miettinen2006} and $\sim 10^{23}$~\cmsq\ respectively, so that
the abundances of \cthree\ are $\sim 10^{-8}$.  It is interesting to
note that warm-up chemical models of the environment around hot cores
similar to the models by \citet{hassel2008},  with
$n$=2\,10$^5$~\cmcub\ and A$_{\rm V}$ = 10  yield an abundance of
4--6$\times 10^{-8}$.

Based on absorption studies of \cthree\ in optical wavelengths the
\cthree\ column densities in diffuse and translucent clouds are found
to range between 10$^{12}$--10$^{13}$~\cmsq\
\citep{maier2001,roueff2002,oka2003}. The \cthree\ column density
observed in the present study is larger by about a factor of 100 or
more than those from the optical studies.  Thus, the non-detection of
\cthree\ in the foreground diffuse gas in the direction of our sources
is consistent with the sensitivity of our observations.

{}

\begin{appendix}
\section{Formulae used for the excitation analysis}

In the approximation of weak absorption ($\tau \ll 1$) and a negligible
population in the upper, $\nu_2=1$, state, the lower state column density is
given by: 

\begin{equation}
N_l = \frac{8\pi \nu^3}{c^3}\frac{g_l}{A_{ul}\,g_u}\int{\tau dv}
\end{equation}

The rotational temperature, \Trot, is calculated from the state specific
column densities by 
\begin{equation}
T_{\rm rot} (J,J') = \frac{E_J-E_{J^\prime}}{k}
\left[\ln
\left(\frac{N_{J^\prime}/(2J^\prime+1)}{N_J/(2J+1)}\right)\right]^{-1}
\end{equation}

where the energy of the levels is given by $E_J = hBJ(J+1)$ and the
rotational constant for the lower vibrational state ( $\nu_2$ = 0) is B
= 12908.242 MHz. 

Note that a ratio of level populations close to unity,
$\left(\frac{N_{J'}/(2J'+1)}{N_J/(2J+1)}\right)\approx 1$, implies
high values of \Trot\ in comparison to the rotational energy scale
defined by $hB/k_{\rm B}$, i.e. 0.62~K in the case of \cthree.  The
formal errors derived for the \Trot\ are correspondingly very large.

Assuming a thermalized population across the rotational ladder with a
unique value of $T_{\rm rot}$ for all levels, a measured single state
column density can be converted to the total column density \Ntot\
using:

\begin{equationarray}{rrc}
N_{C_3} &=& P(T_{\rm rot}) \,\frac{N_J}{2J+1}\, exp\left(\frac{hB}{k_B T_{rot}} J(J+1)\right) \\
	&\approx & \frac{T_{rot}}{2 hB/k_B}\, \frac{N_J}{2J+1}\, exp\left(\frac{hB}{k_B T_{rot}} J(J+1)\right)
\end{equationarray}

where the approximation $T_{\rm rot}\gg hB/k$ 
has been applied for the 
partition function 

\begin{equation}
P(T_{\rm rot})=\sum_{J=0,2,4,\ldots}
(2J+1) \exp \left(-E_J/kT_{\rm rot}\right) \approx \frac{k_{\rm B} T_{\rm
rot}}{2 h B},
\end{equation}

which is half of the usual value for a linear rotor due to the symmetry
not allowing odd-J states.

\end{appendix}

\acknowledgement
HIFI has been designed and built by a consortium
of institutes and university departments from across Europe, Canada
and the United States under the leadership of SRON Netherlands
Institute for Space Research, Groningen, The Netherlands and with
major contributions from Germany, France and the US. Consortium
members are: Canada: CSA, U.~Waterloo; France: CESR, LAB, LERMA, IRAM;
Germany: KOSMA, MPIfR, MPS; Ireland, NUI Maynooth; Italy: ASI,
IFSI-INAF, Osservatorio Astrofisico di Arcetri-INAF; Netherlands:
SRON, TUD; Poland: CAMK, CBK; Spain: Observatorio Astron\'omico
Nacional (IGN), Centro de Astrobiolog\'a (CSIC-INTA).  Sweden:
Chalmers University of Technology - MC2, RSS \& GARD; Onsala Space
Observatory; Swedish National Space Board, Stockholm University -
Stockholm Observatory; Switzerland: ETH Zurich, FHNW; USA: Caltech,
JPL, NHSC. 
JC and JRG thanks spanish MICINN for funding support under projects
AYA2009-07304 and CSD2009-00038. M.S. acknowledge support from grant N
203 393334 from Polish MNiSW.

\end{document}